# Geodesic distance on optimally regularized functional connectomes uncovers individual fingerprints

**Kausar Abbas[1,2], Mintao Liu[1,2], Manasij Venkatesh[3], Enrico Amico[4,5], Alan David Kaplan[6], Mario Ventresca[2], Luiz Pessoa[3,7], Jaroslaw Harezlak[8*], Joaquín Goñi[1,2,9*]**

[1] Purdue Institute for Integrative Neuroscience, Purdue University, West Lafayette, IN, USA

[2] School of Industrial Engineering, Purdue University, West Lafayette, IN, USA

[3] Department of Electrical and Computer Engineering, University of Maryland, College Park, MD, USA

[4] Institute of Bioengineering. Center for Neuroprosthetics. École Polytechnique Fédérale de Lausanne, Lausanne, Switzerland.

[5] Department of Radiology and Medical Informatics, University of Geneva (UNIGE), Geneva, Switzerland

[6] Lawrence Livermore National Laboratory. Livermore, CA, USA

[7] Department of Psychology and Maryland Neuroimaging Center, University of Maryland, College Park, MD, USA

[8] Department of Epidemiology and Biostatistics, Indiana University, IN, USA

[9] Weldon School of Biomedical Engineering, Purdue University, West Lafayette, IN, USA

* both authors contributed equally. corresponding authors: harezlak@iu.edu; jgonicor@purdue.edu

## Abstract

**Background:** Functional connectomes (FCs), have been shown to provide a reproducible individual fingerprint, which has opened the possibility of personalized medicine for neuro/psychiatric disorders. Thus, developing accurate ways to compare FCs is essential to establish associations with behavior and/or cognition at the individual-level.

**Methods:** Canonically, FCs are compared using Pearson's correlation coefficient of the entire functional connectivity profiles. Recently, it has been proposed that the use of *geodesic distance* is a more accurate way of comparing functional connectomes, one which reflects the underlying non-Euclidean geometry of the data. Computing geodesic distance requires FCs to be positive-definite and hence invertible matrices. As this requirement depends on the fMRI scanning length and the parcellation used, it is not always attainable and sometimes a regularization procedure is required.

**Results:** In the present work, we show that regularization is not only an algebraic operation for making FCs invertible, but also that an optimal magnitude of regularization leads to systematically higher fingerprints. We also show evidence that optimal regularization is dataset-dependent, and varies as a function of condition, parcellation, scanning length, and the number of frames used to compute the FCs.

**Discussion:** We demonstrate that a universally fixed regularization does not fully uncover the potential of geodesic distance on individual fingerprinting, and indeed could severely diminish it. Thus, an optimal regularization must be estimated on each dataset to uncover the most differentiable across-subject and reproducible within-subject geodesic distances between FCs. The resulting pairwise geodesic distances at the optimal regularization level constitute a very reliable quantification of differences between subjects.

## Impact Statement

Functional connectomes (FCs) have a reproducible individual fingerprint, making it possible to study neurological and psychiatric phenomena at an individual-level. But this requires an accurate way to compare FCs to establish individual-level associations with behavior and/or cognition. Although the canonical methods of comparing FCs (e.g. correlation, Euclidean) are adequate, geodesic distance provides a more principled and accurate way of comparing FCs by utilizing the underlying non-Euclidean geometry of correlation matrices. We demonstrate that by combining geodesic distance with an optimal amount of regularization, we can get substantially more reliable estimates of relative distances between FCs and thus uncover individual-level differences.

# 1. Introduction

Brain activity can be estimated, indirectly, by measuring the Blood Oxygenation Level Dependent (BOLD) signal using magnetic resonance imaging (MRI) (Bandettini et al., 1992; Frahm et al., 1992; Kwong et al., 1992; Ogawa et al., 1990, 1992). This is the standard technique to generate brain images in functional MRI (fMRI) studies. Functional connectivity between two distinct brain regions is then defined as the statistical dependence between the corresponding BOLD signals, canonically estimated with Pearson's correlation coefficient (Bravais, 1846; Galton, 1886). A whole-brain functional connectivity pattern can be represented as a full symmetric correlation matrix denominated Functional Connectome (FC) (Fornito et al., 2016; Sporns, 2018). FCs have been used to study the changes in brain connectivity with aging (Zuo et al., 2017), cognitive abilities (Shen et al., 2017; Svaldi et al., 2019) and across a wide range of brain disorders (Fornito et al., 2015; Fornito & Bullmore, 2015; van den Heuvel & Sporns, 2019). Recently, it has also been shown that FCs have a recurrent and reproducible individual fingerprint (Abbas et al., 2020; Amico & Goñi, 2018; Finn et al., 2015; Gratton et al., 2018; Mars et al., 2018; Pallarés et al., 2018; Rajapandian et al., 2020; Satterthwaite et al., 2018; Seitzman et al., 2019; Venkatesh et al., 2020), which has opened the possibility of personalized medicine for neuro/psychiatric disorders (Satterthwaite et al., 2018), aided by improved acquisition parameters and the availability of large datasets with open data policy (N. E. Allen et al., 2014; Amunts et al., 2016; Miller et al., 2016; Okano et al., 2015; Poo et al., 2016; D. C. Van Essen et al., 2012; David C. Van Essen et al., 2013).

A clinically useful individual-level biomarker must have high inter-individual differentiability which in turn requires an accurate way of comparing individual FCs. FCs are compared traditionally by computing the Pearson's correlation coefficient between their upper-triangular vectorized versions (Amico & Goñi, 2018; Bari et al., 2019; Finn et al., 2015). This approach enables us to assess to what extent it is possible to identify a participant from a large population of participants, a process known as fingerprinting or subject identification. The success rate of subject identification is known as identification rate (Finn et al., 2015), and has been also referred to as participant identification (Venkatesh et al., 2020). Although comparing FCs using Pearson's correlation coefficient is intuitive and computationally simple, it ignores the underlying geometry of the correlation-based FCs (Venkatesh et al., 2020) and hence has had only limited success in terms of identification rates (Finn et al., 2015).

A geometry-aware approach (Venkatesh et al., 2020) has recently been introduced to establish a more accurate way of measuring distance between any two FCs. FCs computed using Pearson's correlation coefficient between BOLD signals of all brain regions, are objects that lie on or inside a non-linear surface or manifold called the positive semidefinite cone (Figure 1). This non-Euclidean geometry of FCs suggests that the distances between FCs are better measured along a geodesic of the cone. This contrasts with using correlation which is equivalent to the cosine of the angle between demeaned and normalized FCs, or the Euclidean distance which is equivalent to the straight-line distance between FCs. Venkatesh et al. applied the geodesic approach of comparison to the problem of individual fingerprinting and showed that it improves identification rates robustly compared to a dissimilarity measure based on Pearson's correlation coefficient. The improvement was observed across most conditions (resting-state and 7 fMRI tasks) from the Human Connectome Project (HCP) dataset.

The non-optimality of conventional metrics to compare FCs can be shown in another way. When comparing Functional Connectomes (FCs) using the conventional Pearson or Spearman-based correlations, the FCs are vectorized and then correlated. Implicit in this process is the assumption that all the elements of FCs are uncorrelated features. This is not the case. Since FCs are correlation matrices ($Q$), they live on or inside a positive semidefinite cone i.e. $y^T Q y \geq 0$ for all non-zero vectors $y$ (Pennec et al., 2006). This means that elements of $Q$ are inter-related, which violates the implicit uncorrelated feature assumption when conventional metrics are used to compare FCs. Geodesic distance treats $Q$ as a single object, instead of treating each element separately, which results in much more accurate comparisons of FCs evidenced by robustly higher identification rates (Venkatesh et al., 2020).

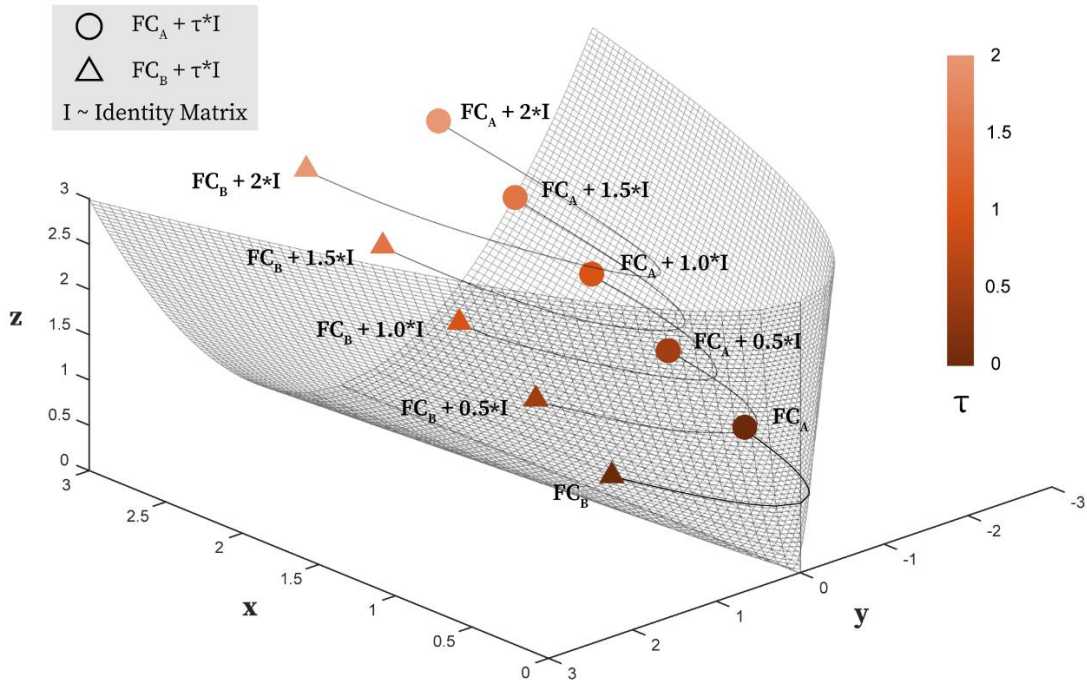

**Figure 1. Incremental regularization of functional connectomes (FCs) and its effect on the estimates of geodesic distance.**
We illustrate the geodesic distance between two FCs of size $2 \times 2$ (denoted here by a circle and a triangle) and how it changes with increasing regularization ($\tau$) of FCs. All the positive-definite (full rank) FCs comprise the cone interior while all the rank-deficient positive semi-definite FCs (having at least one 0 eigenvalue) reside on the cone boundary. Different magnitudes of $\tau$ reallocate FCs within the positive semi-definite cone. We should also highlight that for FCs of higher dimensions, a three-dimensional visualization of the positive semi-definite cone is not possible.

The definition of geodesic distance between two positive definite matrices of the same size (say $Q_1$ and $Q_2$) requires that at least one of the matrices being compared is invertible (Pennec et al., 2006). When this is not the case (rank deficient matrices with at least one eigenvalue equal to 0), both $Q_1$ and $Q_2$ can be regularized by adding a scaled identity matrix, $\tau \times I$, to both, which increases the eigenvalues of both matrices by $\tau$, ensuring that they become invertible. This process was used by Venkatesh et al. (Venkatesh et al., 2020) with a fixed $\tau = 1$, for all fMRI tasks, when computing geodesic distances between (otherwise) rank deficient matrices. Rank deficient FCs may occur typically when the number of time points (from the BOLD time-series) is smaller than the number of brain regions of the parcellation used. It may also happen when using very high resolution parcellations, even if the number of time points are larger than the number of brain regions. Venkatesh et al. demonstrated that even with rank-deficient matrices, with the help of regularization, higher identification rates were achieved using geodesic distance when compared to identification rates based on correlation dissimilarity. Please note that regularization does not affect similarities or dissimilarities between FCs based on Pearson's correlation coefficient as the main diagonal is not even considered for comparison purposes.

Using a regularization of $\tau = 1$ ensures that FCs were invertible and hence permits to use geodesic distance for assessing subject fingerprints as measured by identification rate. However, one could use any positive value of the regularization parameter, $\tau$, and repeat the process of subject identification to assess performance. Geodesic distance is mostly influenced by the eigenvalues of the FCs being compared, which are in turn impacted by the values in their main diagonal. Hence, intuitively, increasing main diagonal regularization is equivalent to shifting and *shrinking* the space occupied by the matrices within the manifold (see Figure 1). Indeed, as $\tau$ tends to infinity all distances between FCs tend towards zero. It can be hypothesized that such regularized shrinking will affect not only absolute but also relative distances between FCs, which in turn would affect identification rates. In such a scenario, full rank and invertible FCs may also benefit from the same regularization procedure to improve identification rates.

In this paper, we explore the effect of the magnitude of the regularization parameter ($\tau$) on the geodesic distance between FCs and its impact on identification rates. We assess this effect for different scanning lengths, number of frames for a fixed scanning length, parcellations and fMRI tasks and evaluate which levels of regularization maximize identification rates. In this manner, we aim to develop a procedure to uncover individual fingerprints by shifting FC data to an optimal location of the semi-definite cone where test/retest FCs are more differentiable across subjects. The existence of an optimal regularization would be critical to obtain the corresponding geodesic distances between FCs. An optimal amount of regularization should lead to higher identification rates in FCs (i.e. higher individual fingerprint), and hence these optimally regularized FCs and particularly their corresponding pairwise distances would be better suited for establishing associations between functional connectivity and cognition, behavior, and neurological diseases at the individual level.

## 2. Methods

### 2.1. Dataset

We included the $N = 426$ unrelated subjects from the Human Connectome Project (HCP) 1,200-participants release (David C. Van Essen et al., 2013). This subset of unrelated subjects was chosen from the overall dataset to ensure that no two subjects have a shared parent. The criterion to exclude siblings (whether they share one or both parents) was crucial to avoid confounding effects in our analyses due to family-structure confounders. Data from resting-state (REST) and seven fMRI tasks were used: emotion processing (EM), gambling (GAM), language (LAN), motor (MOT), relational processing (REL), social cognition (SOC) and working-memory (WM). In this study, we will refer to the resting-state plus all the tasks as *conditions*.

For each condition, subjects underwent two sessions corresponding to two different acquisitions (left to right or LR, and right to left or RL). The resting-state fMRI scans were acquired on two different days with a total of four sessions (coded as REST1 and REST2). The two sessions from REST1 were used for most of the analyses in this study. REST2 sessions were only used in the generalizability analysis (see section 2.6). The HCP scanning protocol was approved by the Institutional Review Board at Washington University in St. Louis. Full details on the HCP dataset have been published previously (Glasser et al., 2013; Smith et al., 2013; D. C. Van Essen et al., 2012).

## 2.2. Brain Parcellations

Two grey matter parcellations were used in this study:

- The **Destrieux** atlas (Destrieux et al., 2010), or "aparc.2009s" in FreeSurfer nomenclature, defined using "Rules and algorithm that produced labels consistent with anatomical rules as well as automated computational parcellation," featuring 75 regions in each hemisphere (74 + Medial Wall), with the particularity of separating gyral and sulcal areas (a total of 150 brain regions).
- **MMP1.0** atlas (Glasser et al., 2016), a multi-modal parcellation of the human cerebral cortex, with 180 brain regions in each hemisphere (a total of 360 brain regions).

For completeness, 14 subcortical regions were added to each parcellation, as provided by the HCP release (filename Atlas_ROI2.nii.gz). To do so, this file was converted from NIFTI to CIFTI format using the HCP workbench software (Glasser et al., 2016; Marcus et al., 2011) (http://www.humanconnecome.org/software/connectome-workbench.html, command -cifti-create-label). This resulted in a total of 164 and 374 brain regions for Destrieux and MMP1.0 parcellations, respectively.

## 2.3. Preprocessing

The data processed using the 'minimal' preprocessing pipeline from the HCP was employed in this work (Glasser et al., 2013). This pipeline included artifact removal, motion correction, and registration to standard template. Full details on this pipeline can be found in earlier publications (Glasser et al., 2013; Smith et al., 2013).

We added the following steps to the 'minimal' processing pipeline. For resting-state fMRI data: (i) we regressed out the global gray-matter signal from the voxel time courses (Power et al., 2014), (ii) we applied a bandpass first-order Butterworth filter in the forward and reverse directions (0.001Hz to 0.08Hz(Power et al., 2014); MATLAB functions *butter* and *filtfilt*), and (iii) the voxel time courses were z-scored and then averaged per brain region, excluding any outlier time points that were outside of 3 standard deviation from the mean (*workbench* software, command *-cifti-parcellate*). For task fMRI data, we applied the same steps as mentioned above but a more liberal frequency range was adopted for the band-pass filter (0.001Hz to 0.25Hz) (Amico et al., 2019), since the relationship between different tasks and optimal frequency ranges is still unclear (Cole et al., 2014).

Table 1 shows the number of frames per run and the scanning length for all fMRI conditions. It also shows the number of participants for whom this number of frames per run were available after the preprocessing. Any runs where we could not fully process the data or were left with fewer frames were left out of the analyses.

**Table 1.** Summary of the number of unrelated participants available (out of a total of 426) for each parcellation and condition after complete preprocessing of the fMRI data with corresponding number of frames per run. Abbreviations ─ REST: resting-state; EM: emotion processing; GAM: gambling; LAN: language; MOT: motor; REL: relational processing; SOC: social; WM: working-memory.

| condition | REST | EM | GAM | LAN | MOT | REL | SOC | WM |
|---|---|---|---|---|---|---|---|---|
| total participants – Destrieux | 407 | 408 | 408 | 409 | 409 | 409 | 409 | 409 |
| total participants – MMP1.0 | 405 | 406 | 406 | 407 | 407 | 407 | 407 | 407 |
| frames per run | 1190 | 166 | 243 | 306 | 274 | 222 | 264 | 395 |
| scanning length (min) | 14.28 | 1.99 | 2.92 | 3.67 | 3.29 | 2.66 | 3.17 | 4.74 |

## 2.4. Whole-brain Functional Connectomes

As described in Section 2.3, for a given brain parcellation, time series data for each voxel was z-scored and averaged within each brain region. Pearson's correlation coefficient (MATLAB command *corr*) was used to estimate the functional connectivity between all pairs of brain regions, resulting in a symmetric correlation matrix of size $m \times m$ where $m$ is the number of brain regions in the parcellation being used. This object is referred to as a Functional Connectome (FC). A whole-brain FC was computed for each of the two runs of each participant and each condition (resting-state and seven tasks).

As mentioned above, FCs are correlation matrices and it is well known that correlation matrices are symmetric positive semi-definite (SPSD), which means their eigenvalues are greater than or equal to zero (Bhatia, 2009). If all the eigenvalues of an FC are strictly greater than zero, then it is a symmetric positive *definite* (SPD) FC matrix. The rank and invertibility of an FC are also directly related to its eigenvalues: if one or more eigenvalues are zero, then that FC is *rank-deficient* and not invertible. When all the eigenvalues are greater than zero for an FC, it is *full-rank* and hence invertible (Bhatia, 2009). The rank of an FC depends on the number of brain regions in the parcellation ($m$) and the number of samples in the time series ($T$) such that:

$$rank \leq m \quad \text{for } T \geq m$$

$$rank < T \quad \text{for } T < m$$

For all the conditions, the FCs generated using Destrieux parcellation were full-rank if the number of samples (frames) in the time series used was $\geq 164$ (e.g. when using entire scanning for any condition) while the FCs generated using MMP1.0 parcellation were always rank-deficient, regardless of the number of samples in the time series (see Table 1).

*2.5. Geometry of Functional Connectomes*

Functional Connectomes (FCs) estimated using Pearson's correlation coefficient are objects that lie on or inside a non-linear surface, or manifold, called the positive semi-definite cone. Although a three-dimensional visualization of this manifold is only possible for $2 \times 2$ FCs (see Figure 1), a manifold with exactly the same properties exists for FCs with higher dimensions (Bhatia, 2009). Pearson's correlation coefficient is the canonical way to estimate similarity/dissimilarity between FCs (Amico & Goñi, 2018; Finn et al., 2015), while other related approaches, such as Euclidean distance between the vectorized matrices (Ponsoda et al., 2017) and the so called Manhattan ($L^1$) distance (E. A. Allen et al., 2014), have also been used. Considering the non-Euclidean geometry of FCs, it is natural to measure the distance between FCs along the curvature of the positive semi-definite cone (Bhatia, 2009). The geodesic distance between two points inside the cone, thus between two SPD FCs $Q_1$ and $Q_2$, is the shortest path between them along the manifold and is unique for any two such points (Bhatia, 2009; Pennec et al., 2006).

Let $\mathbb{S}_+^m$ be the set of all symmetric positive matrices of dimension $M$, which lie on or inside a symmetric positive semi-definite cone of dimension $M$. The positive-definite matrices would comprise the interior of the cone while all the rank-deficient semi-definite matrices would reside on the cone boundary. Now assume that $Q_1 \in \mathbb{S}_+^m$ and $Q_2 \in \mathbb{S}_+^m$ are two SPD matrices of size $m \times m$ (here, $m = 164$ or $374$). Let us denote $Q = Q_1^{\frac{-1}{2}} Q_2 Q_1^{\frac{-1}{2}}$, then $Q \in \mathbb{S}_+^m$ and its corresponding $m$ eigenvalues satisfy $\tau_i \geq 0 (1 \leq i \leq m)$. Then the geodesic distance between $Q_1$ and $Q_2$ is computed as (Bhatia, 2009; Pennec et al., 2006):

$$d_G(Q_1, Q_2) = \sqrt{trace\left(log^2\left(Q_1^{\frac{-1}{2}} Q_2 Q_1^{\frac{-1}{2}}\right)\right)} = \sqrt{\sum_{i=1}^{m}(log(\tau_i))^2} \qquad (Eq.1)$$

where $log$ is the matrix log operator. This definition of geodesic distance requires that the matrix $Q_1$ is invertible (or equivalently SPD or full-rank). When this is not the case, we can regularize both $Q_1$ and $Q_2$ by adding to each of them a scaled identity matrix, $\tau \times I$, which increases the value of their eigenvalues by $\tau$, ensuring that they are now invertible matrices. Importantly, this regularization reallocates both matrices within the positive semi-definite cone (Figure 1).

Venkatesh et al. (Venkatesh et al., 2020) used $\tau = 1$ with the specific purpose of ensuring full rank matrices. However, theoretically, one could use any positive value of the regularization parameter, $\tau$, to ensure that both matrices ($Q_1$ and $Q_2$) are full-rank. As mentioned earlier, all correlation matrices are either positive definite or positive semi-definite, which means that either all their eigenvalues are positive or at least one of them is zero (they cannot have negative eigenvalues). Thus, even a small positive perturbation to a rank-deficient correlation matrix using a scaled identity matrix would make it full-rank and invertible (i.e. all eigenvalues greater than zero).

## 2.6. Subject Identification

Subject identification is the process of identifying an individual's FC from a population of FCs, given another FC of that individual. All conditions (resting-state and seven tasks) in our dataset contain 2 runs (LR and RL acquisition orientation), which we denominate here Test and Retest. In order to avoid any bias due to the acquisition orientation, runs were randomly assigned to either Test or Retest for each subject. This process was repeated for each condition separately.

An FC from the Retest data was labeled with the participant's identity in the Test data that was closest to it in the Test data. We repeated this process for all the FCs in the Retest data and defined the identification rate as:

$$Identification\ Rate = \frac{Number\ of\ correctly\ labeled\ subjects}{Total\ number\ of\ subjects}$$

This process was repeated by reversing the roles of test and retest sessions, as introduced by Finn and colleagues (Finn et al., 2015). The final identification rate was obtained by averaging the two values.

The identification rates were computed for each condition separately. To study the effects of regularization on the identification rates, this process was repeated for a wide range of regularization parameter values, $\tau$, in particular:

*For Destrieux parcellation:* $\tau = \begin{cases} \langle 0 \rangle\ to\ \langle 2 \rangle & in\ steps\ of\ 0.1 \\ \langle 2.5 \rangle\ to\ \langle 10 \rangle & in\ steps\ of\ 0.5 \end{cases}$

*For MMP1.0 parcellation:* $\tau = \begin{cases} \langle 0 \rangle\ to\ \langle 10 \rangle & in\ steps\ of\ 0.5 \\ \langle 11 \rangle\ to\ \langle 20 \rangle & in\ steps\ of\ 1 \end{cases}$

Different values of $\tau$ for the two parcellations were chosen based on preliminary exploration of the change in identification rates with $\tau$.

To understand the effect of scanning length, for each value of $\tau$, the identification process was repeated by selecting frames *sequentially* out of the total time series, starting from 50 frames to the maximum number of frames, in steps of 50 (see Table 1 for maximum number of available frames and the corresponding scanning length for all eight fMRI conditions).

To understand the effect of number of frames when the scanning length is fixed, the identification process was repeated for each value of $\tau$ using the maximum scanning length. The number of frames was adjusted by choosing alternating frames from the time series i.e. by picking every $2^{nd}$, $3^{rd}$, $4^{th}$ … frame. Note that this process is equivalent to assessing identification rates for longer repetition-times (TRs). The maximum gap between chosen frames was decided for each condition to keep at least 50 frames in the final time series.

To assess variability in identification performance due to differences in samples, we used sampling without replacement. For every run, we randomly selected 80% of the participants and performed subject identification process. This procedure was repeated 100 times for each value of $\tau$ and for each number of frames evaluated.

The above mentioned 'sampling without replacement' process would also serve as a proxy exploration of the generalizability of the optimal regularization magnitude for outside datasets of same or similar acquisition parameters as the ones used in this study. To explore generalizability of the optimal regularization magnitude across different sessions of the same subjects, two sessions from REST2 were used to compare the identification rates for varying values of $\tau$ with REST1, using the entire scanning length.

## 3. Results

We explored the effect of using different values of the regularization parameter ($\tau$) on the geodesic distance, and the uncovering of individual fingerprint in FCs. Identification rate (Finn et al., 2015; Venkatesh et al., 2020) was used as a metric to quantify the individual fingerprint. Identification rate was computed by the Subject Identification process, which is the process of identifying an individual's FC from a population of FCs, given another FC of that individual. Identification rate is simply the percentage of accurately identified individuals. Through a small example, we show evidence of regularization affecting not only the global geodesic distance but also relative distance between FCs, which ultimately may affect identification rates. Then we systematically studied how regularization affects identification rates for FCs, with different fMRI conditions (resting-state and seven fMRI tasks), parcellations, varying scanning lengths, and finally, varying number of frames for a fixed scanning length. The generalizability of the optimal regularization magnitude for different sessions of the same subjects and for different subjects for whom fMRI data was acquired with exactly same acquisition parameters, was also investigated.

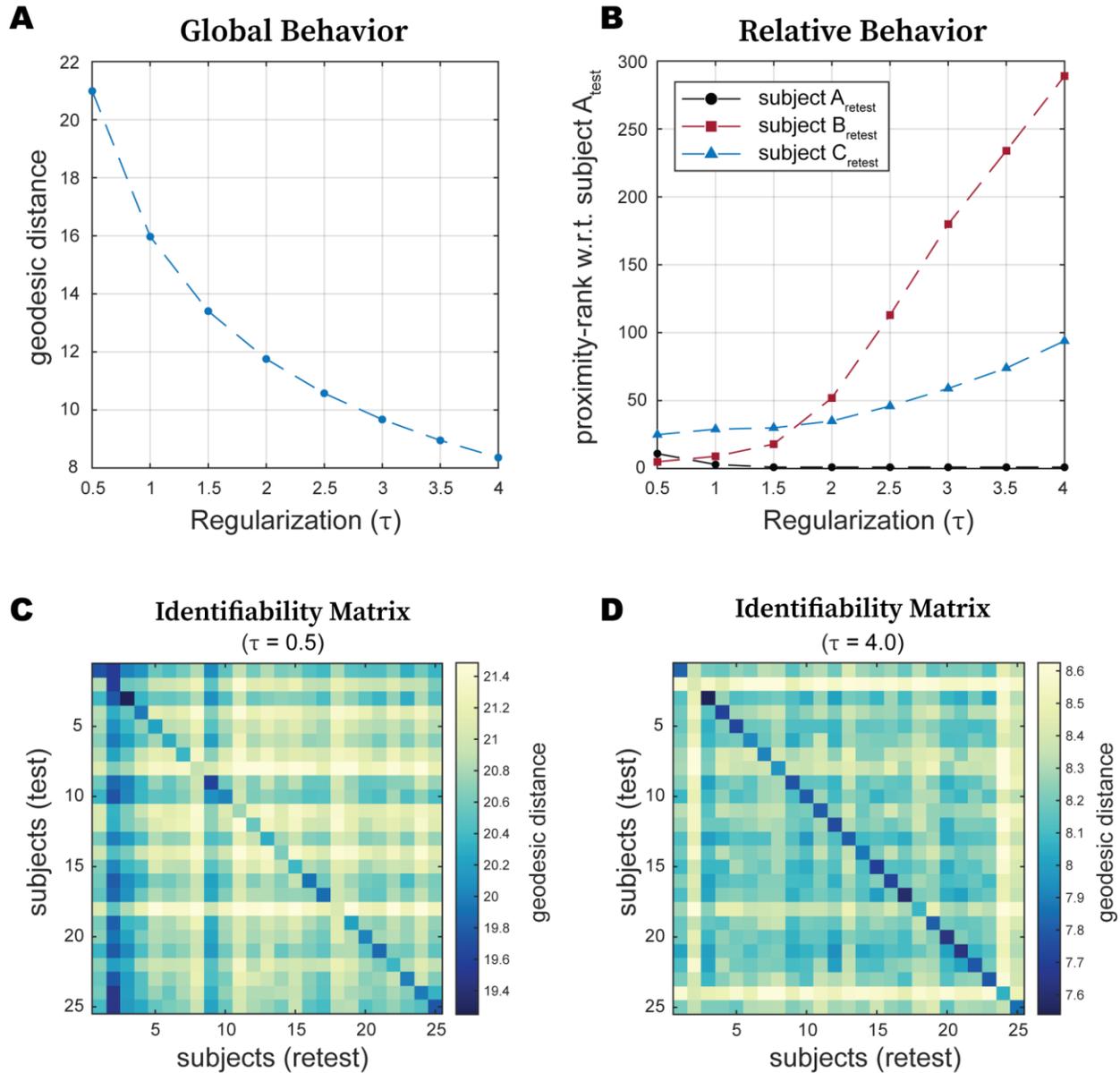

Figure 2: **Effect of regularization ($\tau$) on global and relative geodesic distances.** We have chosen the emotion processing FCs to illustrate how geodesic distances across subjects and/or sessions change with regularization magnitude. **(A)** Global geodesic distance (in this case averaged geodesic distance between test and retest sessions of all subjects for the emotion processing task) decreases exponentially with increasing regularization. **(B)** shows how close (in terms of proximity-rank with respect to all the other subjects) retest sessions of subjects A, B and C are to the test session of subject A. Note that the three proximity-ranks fluctuate with regularization. **(C)** Identifiability matrix based on geodesic distance for low ($\tau = 0.5$) regularization for a subsample of 25 subjects performing the emotion processing task. **(D)** Identifiability matrix based on geodesic distance for high ($\tau = 4$) regularization for the same subsample of 25 subjects performing the emotion processing task.

We first provided an example to develop an intuitive understanding of how regularization affects geodesic distances between FCs. To do so, we assessed the effect of regularization on geodesic distances among FCs when subjects are performing the emotion processing task. Figure 2A shows that as regularization ($\tau$) increases, average geodesic distance across all subjects and sessions (global geodesic distance), exponentially decreases. We then assessed the effect of regularization on the relative geodesic distances between FCs. Figure 2B shows the proximity-rank in terms of distance. Briefly, the proximity-rank of an FC B with respect to an FC A quantifies how many FCs in that dataset are closer to FC A than FC B. Taking as reference subject $A_{test}$, we tracked the proximity-rank of the subjects $A_{retest}$, $B_{retest}$, and $C_{retest}$ at different levels of $\tau$. At $\tau = 0.5$, compared to $B_{retest}$, there are many more FCs closer to $A_{test}$ than $C_{retest}$. However, the situation is reversed at $\tau = 4$. In addition, at $\tau = 0.5$, $B_{retest}$ is closer $A_{test}$ than $A_{retest}$. Last but not least, the proximity-rank of the $A_{retest}$ also changes with respect to $A_{test}$ with regularization.

Results above show that regularization not only affects the global geodesic distance among FCs, but also the relative distance, which may ultimately affect identification rates. Figure 2C−D show the identifiability matrices for 25 subjects chosen arbitrarily (for ease of visualization) performing the emotion processing task at a low ($\tau = 0.5$) and a high ($\tau = 4$) value of regularization. An $(i, j)$ entry in an identifiability matrix here shows the geodesic distance between the test FC of the *i-th* and the restest FC of the *j-th* subject. Since the order of the subjects is the same across rows and columns, a main diagonal entry represents the geodesic distance between test and re-test session of the same subject. Thus, a brighter main diagonal in Figure 2D, compared to 2C, indicates that the test and retest sessions of the same subject are closer to each other than other subjects at $\tau = 4$ than at $\tau = 0.5$, which should translate into higher identification rates.

Intuitively, these results tell us that asymptotically, geodesic distances between FCs approach zero as $\tau$ tends to infinity. In addition to affecting the absolute magnitude of the distances, $\tau$ also affects the relative distances between FCs and we have preliminary evidence that there is an optimal value/range of $\tau$ which would affect relative distances in such a way that FCs from the two sessions of the same subject are closer to each other than any other FCs. These findings motivate us to assess changes in subject identification rates with varying magnitudes of $\tau$.

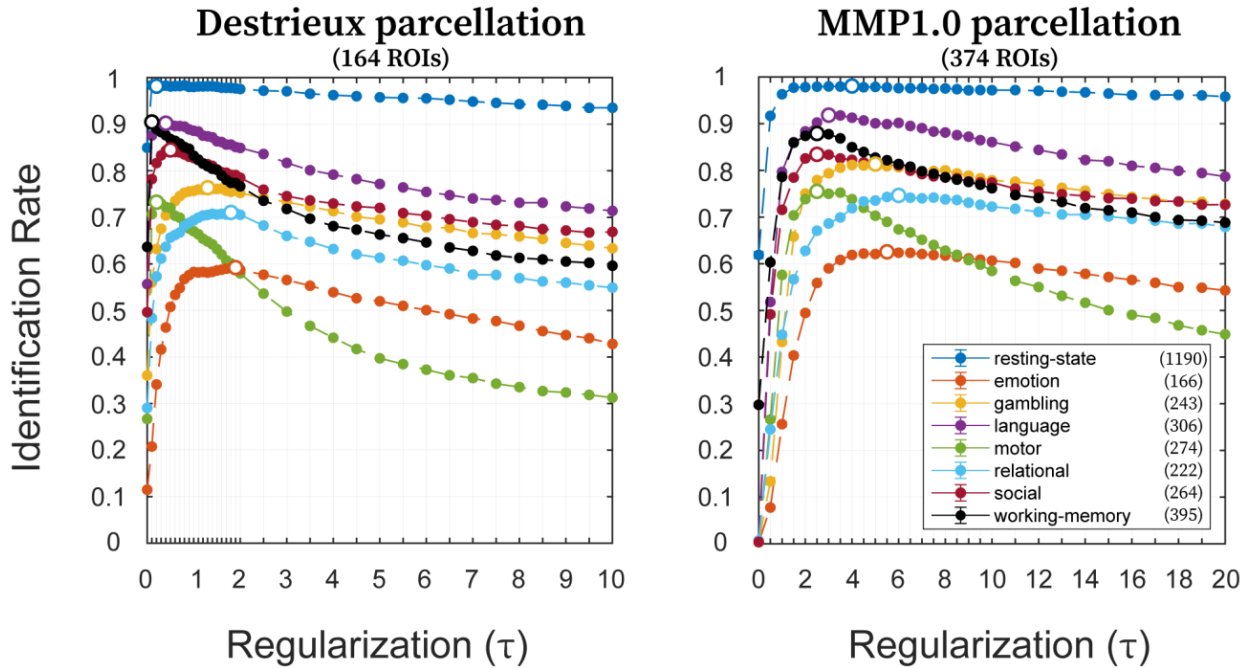

**Figure 3: Effect of regularization ($\tau$) on identification rates.** Identification rates for all eight conditions (utilizing maximum available scanning length) with variable magnitudes of $\tau$, using Destrieux (left; 164 ROIs) and MMP1.0 (right; 374 ROIs) parcellations. Filled circles indicate the mean identification rate while error bars indicate the standard error of the mean across samplings with replacement (error bars are small enough that they are hidden behind the circles). Legend indicates the eight conditions along with maximum available number of frames. Along each curve, the circle not filled indicates the optimal value of $\tau$ which maximizes the identification rate.

Figure 3 shows the effect of $\tau$ on identification rates for all fMRI conditions (using the entire scanning length) and for both the Destrieux and MMP1.0 parcellations. Identification rates for all conditions and different parcellations appeared to be highly sensitive and roughly concave functions of $\tau$. In most cases, we observed the presence of a clearly identifiable optimal $\tau$ (from now on denominated $\tau^*$) value for which the identification rate is maximized. For a few cases for the MMP1.0 parcellation, it seems that there was a wide range of optimal $\tau$ that produced very similar identification rates (e.g. resting-state, emotion).

**Table 2.** Optimal identification (ID) rates for all eight fMRI conditions using **Destrieux** parcellation, and the corresponding values of the optimal scan length, percentage of maximum available frames, and the optimal regularization magnitude ($\tau^*$). Abbreviations – REST: resting-state; EM: emotion processing; GAM: gambling; LAN: language; MOT: motor; REL: relational processing; SOC: social; WM: working-memory.

| condition | REST | EM | GAM | LAN | MOT | REL | SOC | WM |
|---|---|---|---|---|---|---|---|---|
| optimal ID rate | 0.98 | 0.59 | 0.76 | 0.90 | 0.73 | 0.71 | 0.84 | 0.90 |
| optimal scan length (min: sec) | 13:48 | 1:59 | 2:55 | 3:36 | 3:17 | 2:40 | 3:10 | 4:44 |
| % of frames | 97 | 100 | 100 | 98 | 100 | 100 | 100 | 100 |
| $\tau^*$ | 0.2 | 1.9 | 1.3 | 0.5 | 0.2 | 1.8 | 0.5 | 0.1 |

Using the entire scanning length, $\tau^*$ depended not only on the condition but also on the parcellation (Figure 3). The $\tau^*$ values were smaller for the Destrieux parcellation than for the MMP1.0 parcellation for any given condition. Resting-state, language and working-memory had the highest, while the emotion task had the lowest identification rates at $\tau^*$ for both parcellations. At $\tau^*$, the identification rates were either approximately equal (for resting-state) or higher when using MMP1.0 parcellation, compared to Destrieux, except for working-memory and social tasks. For both parcellations, resting-state condition reached greater than 99% identification rate at $\tau^*$.

**Table 3.** Optimal identification (ID) rates for all eight fMRI conditions using **MMP1.0** parcellation, and the corresponding values of the optimal scan length, percentage of maximum available frames, and the optimal regularization magnitude ($\tau^*$). Abbreviations – REST: resting-state; EM: emotion processing; GAM: gambling; LAN: language; MOT: motor; REL: relational processing; SOC: social; WM: working-memory.

| condition | REST | EM | GAM | LAN | MOT | REL | SOC | WM |
|---|---|---|---|---|---|---|---|---|
| optimal ID rate | 0.98 | 0.63 | 0.81 | 0.92 | 0.76 | 0.75 | 0.83 | 0.88 |
| optimal scan length (min: sec) | 14:17 | 1:59 | 2:55 | 3:40 | 3:17 | 2:40 | 3:10 | 4:44 |
| % of frames | 100 | 100 | 100 | 100 | 100 | 100 | 100 | 100 |
| $\tau^*$ | 4.0 | 5.5 | 5.0 | 3.0 | 2.5 | 6.0 | 2.5 | 2.5 |

We then assessed the effect of scanning length on identification rate and how it interacts with $\tau$. Results are shown in Figure 4 (Destrieux) and Figure 5 (MMP1.0). With the Destrieux parcellation

(164 brain regions), in general, $\tau^*$ was particularly small (0.1–0.2) for resting-state compared to most tasks (0.2–1.9), with identification rates decreasing slowly with increasing magnitudes (see Figure 4). Overall, for resting-state compared to tasks, the scanning length used to compute FCs played a much bigger role in identification rates than the regularization magnitude. For a given $\tau$, the identification rates tended to increase with increasing scanning length for all conditions, with maximal identification rates achieved with entire scanning length (see Table 2; resting-state and language conditions being the exceptions). With shorter scanning lengths, a broader range of $\tau$ optimized the identification rates. But as the scanning length increased, this range became narrower and hence maximal identification rates required more specific regularization. Also, the drop off in identification rates was sharper when the optimal $\tau$ range became narrower. This pattern is less clear with the emotion processing task, perhaps due to emotion-processing being the fMRI condition with the shortest scanning length than the other tasks (see Table 1).

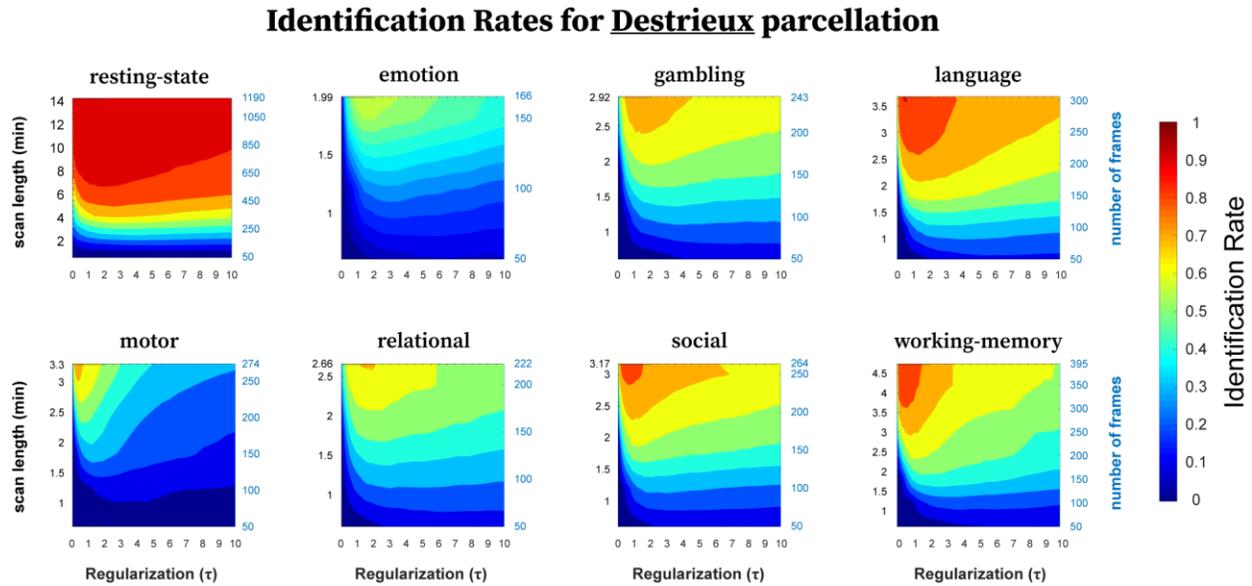

**Figure 4: Identification rates as a function of regularization ($\tau$) and scanning length used to compute FCs using Destrieux parcellation.** The panel shows identification rates, averaged across samplings without replacement, for all eight fMRI conditions. For any given condition, the scanning length was adjusted by selecting frames *sequentially* out of the total time series ranging from 50 to maximum number of frames available, in steps of 50.

With MMP1.0 parcellation (374 brain regions), we observed similar results. Just as with the Destrieux parcellation, resting-state behaved differently than tasks. First, for any given scanning length, τ* values were much smaller for resting-state than for tasks (Figure 5). Second, the identification rates for resting-state were more dependent on the scanning length than on the regularization. For a given τ, identification rates tended to increase with increasing scanning length for all conditions, with maximal identification rates achieved with entire scanning length (Table 3). Finally, the optimal ranges of τ were broader with shorter scanning length and more specific with increasing scanning length. In comparison to the Destrieux parcellation, the narrowing of the optimal τ range required longer scanning length for MMP1.0 for any given condition.

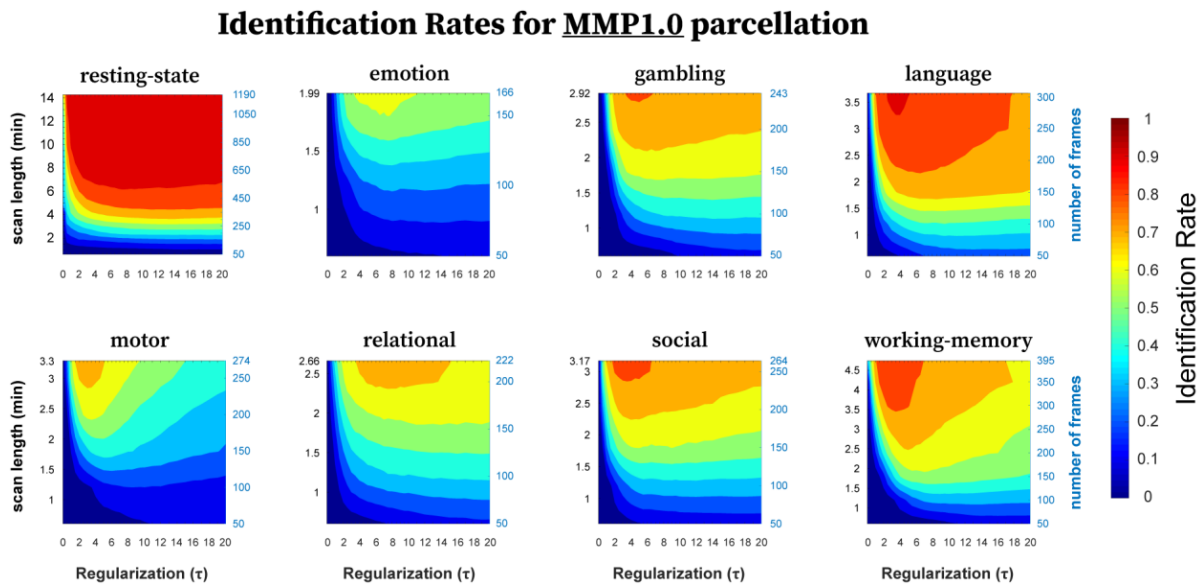

**Figure 5: Identification rates as a function of regularization (τ) and scanning length used to compute FCs using MMP1.0 parcellation.** The panel shows identification rates, averaged across samplings without replacement, for all eight fMRI conditions. For any given condition, the scanning length was adjusted by selecting frames *sequentially* out of the total time series ranging from 50 to maximum number of frames available, in steps of 50.

We also assessed the effect of number of frames on the identification rates, when maintaining the entire scanning length. Overall, for a given condition, the identification rate was not severely affected by decreasing the number of frames (Figure 6). When the number of frames became too small (different for each condition), identification rates dropped more drastically for the Destrieux parcellation than for the MMP1.0. It is interesting to note that with approximately 170 or more frames, identification rates reach a plateau for all fMRI conditions and parcellations.

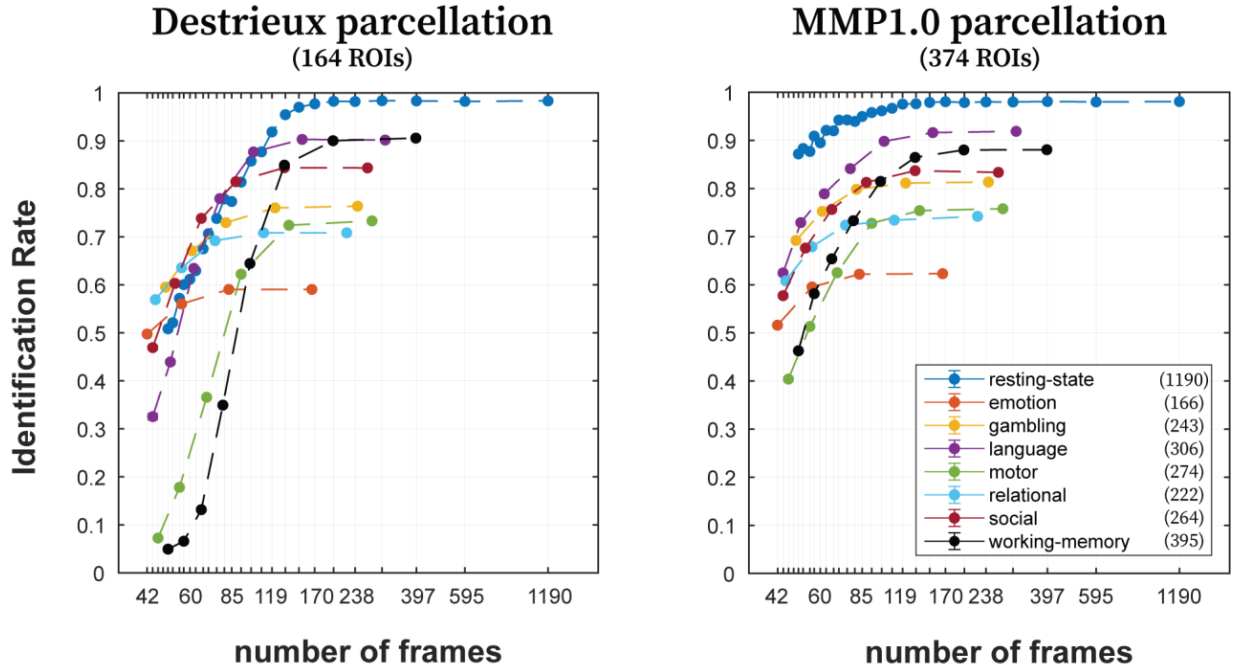

**Figure 6: Effect of number of frames on identification rates using the entire scanning length.** Identification rates for all eight fMRI conditions (utilizing optimal regularization magnitude ($\tau^*$) ─ see Table 1) with variable number of frames, using Destrieux (left; 164 ROIs) and MMP1.0 (right; 374 ROIs) parcellations. Maximum scanning length was always maintained for each condition by choosing alternate points from BOLD time series. For instance, 397 frames were obtained for resting-state by choosing every third time point. Filled circles indicate the mean identification rate while error bars indicate the standard error of the mean across samplings with replacement (error bars are small enough that they are hidden behind the dots). Legend indicates the eight fMRI conditions along with total scanning length.

A very low standard error of mean (SEM) was observed for all the analyses discussed above (Figure 3−6), highlighting the generalizability of the optimal regularization magnitude to FCs from different subjects. Optimal regularization magnitude and the corresponding identification rates for REST2 were found to be similar to REST1 (Figure 7) highlighting the generalizability across different sessions of the same subjects. It should be noted that for both REST1 (Figure 3) and REST2 (Figure 7), there is a range of $\tau$ where the corresponding identification rates are approximately equal to the optimal identification rate. In addition, the scatter plots between identification rates of REST1 and REST2 show how similarly the two samples behave with respect to $\tau$ (Figure 7; insets).

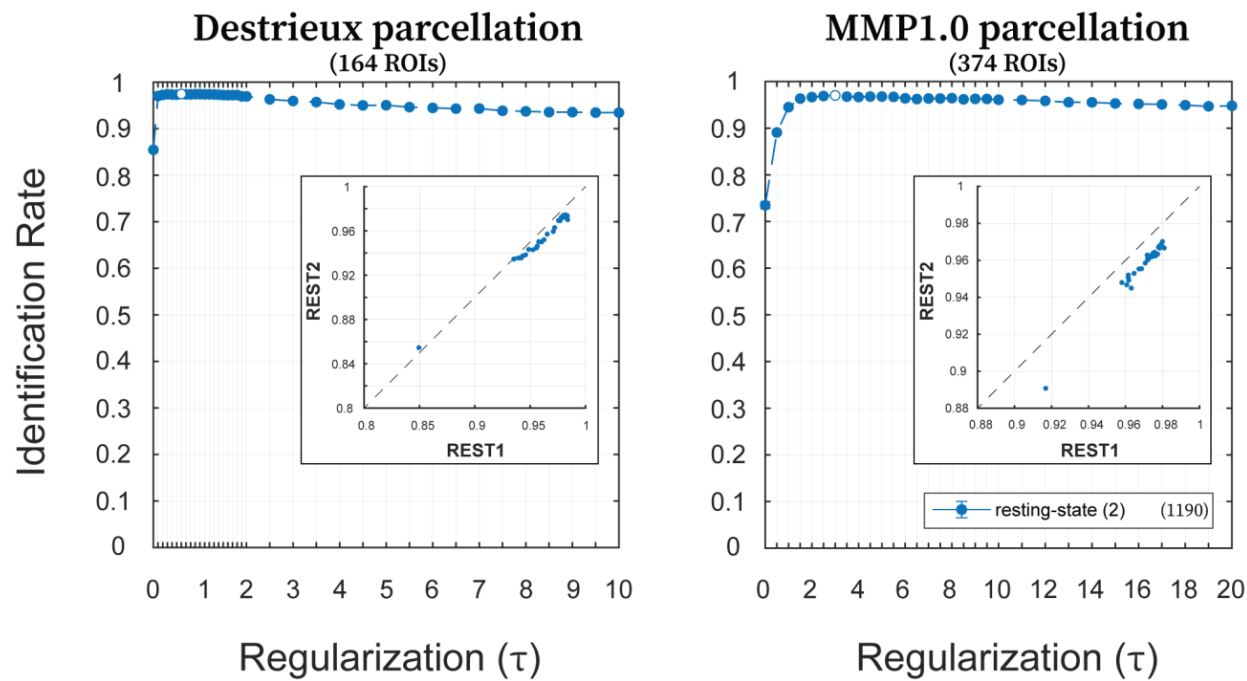

**Figure 7: Generalizability: effect of regularization ($\tau$) on identification rates for REST2.** Identification rates for the two sessions (LR and RL) from REST2 (utilizing maximum available scanning length) with variable magnitudes of $\tau$, using Destrieux (left; 164 ROIs) and MMP1.0 (right; 374 ROIs) parcellations. Filled circles indicate the mean identification rate while error bars indicate the standard error of the mean across samplings with replacement (error bars are small enough that they are hidden behind the circles). Legend indicates the REST2 condition along with maximum available number of frames. Along each curve, the circle not filled indicates the optimal value of $\tau$ which maximizes the identification rate. The insets in both plots are the scatter plots between REST1 and REST2 of the mean identification rates (across samplings) for the entire range of $\tau$. Both *x*- and *y*-axes indicate identification rates and the dotted line is identity line.

## 4. Discussion

In this work, we explored the effects of different magnitudes of regularization on geodesic distance and subsequently its impact on subject identification rates in Functional Connectomes (FCs). We explored these effects for eight fMRI conditions from the HCP data ─ resting-state, emotion, gambling, language, motor, relational, social and working-memory. We found that the optimal value of the regularization parameter, which maximized the identification rates, is dependent on the condition, parcellation, scanning length and the number of frames used to the compute the FCs. In addition, the deviation from the optimal point could affect the identification rates drastically depending on the condition, scanning length and/or the number of frames used. We also found that the magnitude of optimal regularization is generalizable across different subjects and different sessions of the same subjects, when the acquisition parameters are the same. In short, we found that geodesic distance, which has been shown to be a more accurate way of comparing FCs than canonical methods (Venkatesh et al., 2020), can be further refined by choosing an optimal regularization magnitude for each dataset and fMRI condition.

### *4.1. Increased regularization reduces geodesic distance globally and alters relative distances between FCs.*

Geodesic distance is highly determined by the eigenvalues of the FCs being compared (see Eq. 1). When those FCs are regularized by adding a constant value to their main diagonal, it increases their eigenvalues by the same amount, thus affecting the geodesic distance between them. As the regularization magnitude increases, the eigenvalues of the FCs, and hence the geodesic distance between them, becomes dominated by it. Since the regularization value added to both FCs is always equal, for a large enough regularization magnitude, their eigenvalues also become approximately numerically equal, leading to a decreased geodesic distance. Intuitively, increasing main diagonal regularization is equivalent to shifting and *shrinking* the space occupied by the matrices within the manifold. Thus, as the regularization magnitude increases, it was expected that the geodesic distance between FCs would decrease, as observed in Figure 2A.

It was less intuitive that the relative magnitude of the distances would also change with regularization. As the regularization magnitude increased, the relative distance between FCs changed in different directions as shown in Figure 2B with FC $B_{retest}$ and $C_{retest}$. Furthermore, for an optimal value of regularization, the distances between sessions of the same subjects became smaller than between subjects, which lead to better identification of the subjects when comparing the test and the retest sessions, as shown in Figure 2C−D identifiability matrices.

Overall, we can think of increasing regularization as a non-linear shrinking procedure which does not preserve relative distances between FCs. By tracking the effects of regularization on three subjects, we demonstrated that the local distance information is not preserved for different magnitudes of regularization (Figure 2B). This result must be taken into account when using geodesic distance to compare FCs. Then the question is how to decide what magnitude of regularization to choose? The answer lies in the implicit hypothesis that the FCs from two sessions of the same subject should be closer to each other than FCs from any session of any other subject. If we can find a regularization magnitude where for most subjects, this statement is true, then that's the spot where the distances between FCs are the most meaningful, if not accurate. This *optimal* spot can be discovered by tracking identification rates as they change with regularization, as was done in this study.

## *4.2. Identification rate is a concave function of the regularization parameter*

We observed that for any condition and parcellation, there was a specific value or a range of values for the regularization parameter where identification rate peaked (Figure 3). In other words, identification rate was a concave function of the regularization parameter for all fMRI conditions and parcellations tested here. We should emphasize that only a limited range of the regularization parameter was tested in this study, for specific conditions and parcellations, and thus we cannot theoretically guarantee that the optimal levels of regularization found here could be trivially extrapolated to other datasets with different acquisition parameters. But, considering the breadth of the fMRI conditions and the size of the dataset used in this study, we are confident this concave behavior would be replicable in other fMRI datasets as well.

### 4.3. Optimal regularization parameter depends on the specific dataset

We observed that the optimal value of the regularization parameter, which maximizes the identification rates, depends on the condition, parcellation, scanning length, and number of frames used to compute the FCs (Figure 4−5; Table 2−3). Venkatesh et al. (Venkatesh et al., 2020) used a fixed regularization magnitude ($\tau = 1$). Here we show that identification rates can be substantially improved by using dataset-dependent regularization parameter. In addition, although Venkatesh et al. (Venkatesh et al., 2020) employed regularization only when the FCs being compared were rank-deficient, we found that regularization improves identification rates even with full-rank FCs; e.g. this was the case of Destrieux parcellation when $\geq 164$ frames were used to compute FCs.

### 4.4. Longer scanning length leads to more specific values of optimal regularization and to higher identification rates

As the number of samples (or frames chosen sequentially), and hence the scanning length, increase in the time series data, the resultant correlations become more reliable (Bonett & Wright, 2000) and thus we get better estimates of FCs in the 'static' sense of functional connectivity. For all the tasks, we observed that as the scanning length increased, the range of values of $\tau$ which resulted in maximized identification rates narrowed down (Figure 4−5). This effect was not as prominent in resting-state, where for most of the scanning lengths evaluated, there was a wide range of values of $\tau$ which resulted in maximum identification rates. This suggests that resting-state FCs, in comparison to tasks, may reside in an intrinsically different region of the semi-definite cone where reallocation of FCs through regularization does not have a sizeable influence on their differentiability.

It should also be pointed out that with optimal values of $\tau$, the optimal identification rates were almost always obtained when using the entire scanning length (two exceptions: resting-state and language using Destrieux parcellation; Table 2−3). Even in the two cases where it wasn't, the optimal scanning length was marginally smaller than the entire scanning length and the optimal identification rate was approximately equal to the identification rate obtained with maximum scanning length (within margin of error). Intuitively, we can say that the longer the scan acquired, the more information we have about the condition and the subject, which results in higher identification rates.

*4.5 Number of frames and TR length are not as influential as scanning length*

For all conditions, across the two parcellations, when the scanning length was decreased, the identification rates dropped, sometimes drastically (Figure 4−5). Ostensibly, it might seem that this does not hold for resting-state condition but it is worth noting that resting-state scan is a considerably longer acquisition (14 min and 47 sec compared to second longest, working-memory which is 4 min and 44 sec) than all the tasks and the effect of shorter scanning length comes into play when the reduced scanning length becomes comparable to tasks (around 6−7min). The decrease in identification rate with decreasing scanning length raises a natural question: what would happen if scanning length is maintained but the number of frames is reduced?

The answer is that identification rates are considerably less sensitive to number of frames than the scanning length, when the number of frames is not too small (Figure 6). To achieve fewer number of frames while maintain the scanning length, we chose alternate time points, with varying gaps, which introduced another variable into the mix: repetition time i.e. TR. For instance, by choosing every 4th sample from a time series, we are effectively increasing the TR four-fold. So, another conclusion that we could draw from this result is that identification rates are considerably less sensitive to TR length than scanning length. This effect has been observed before by Horien et al. (Horien et al., 2018) but using Pearson's correlation coefficient as a metric to compare FCs. This knowledge could be helpful in designing scanning protocols where often one has to 'sacrifice' spatial resolution for temporal resolution or vice versa. Knowing that as long as one has a long enough scan, perhaps a relatively longer TR could be acceptable in favor of improved spatial resolution, without any detrimental effects to the FC fingerprint.

### *4.6. Regularization counteracts the effect of a coarser grain parcellation on individual fingerprint*

Using Pearson's correlation as a similarity metric to compare FCs, Finn et al. (Finn et al., 2015) showed that a parcellation with more ROIs resulted in higher subject identification rates than a parcellation with fewer ROIs. Venkatesh et al. (Venkatesh et al., 2020) observed the same trend with both geodesic distance and Pearson's correlation-based dissimilarity. This suggested that finer parcellations lead to more uniqueness or fingerprint, at least up to a certain resolution. In this work, we found that when using a coarser resolution parcellation, we can achieve similar identification rates than a finer resolution parcellation when applying geodesic distance with optimal regularization magnitude.

When computing FCs, an ROI time series is computed by averaging voxel-level time series for all the voxels contained within the ROI. One of the main reasons this is done is to increase the signal to noise ratio of the time series under consideration, as the voxel-level time series would be much noisier than an averaged ROI time series. By choosing a finer resolution parcellation, we chose smaller size ROIs, and hence compromise on the signal to noise ratio in the time series in favor of spatial resolution, compared to a coarse resolution parcellation, where an ROI time series would be computed by averaging over a larger number of voxels. Since by using geodesic distance with optimal regularization, we can overcome the downside of coarse resolution parcellation in terms of fingerprint, perhaps we can favor a relatively coarser parcellation for an improved signal to noise ratio while maintaining the individual fingerprint.

### *4.7. Generalizability of the optimal regularization magnitude*

Very small differences were observed in the optimal identification rates and the optimal magnitudes of the regularization parameter when different sub-samples of the dataset were used for subject identification (Figure 3−6). This highlights that the results for optimal regularization in this study are generalizable to other datasets, given that the scans are acquired with the same or similar parameters. If one is to change the acquisition parameters though, the optimal regularization magnitudes might be different. Using the two sessions from REST2 (not used in any of the former analyses), we were also able to show that the optimal regularization magnitudes and the corresponding identification rates are generalizable to different sessions of the same subjects, even when acquired on different days with the same parameters (Figure 7). In addition, we observed that optimal identification rates are maintained for the same amount of regularization when the TR length is increased (to a certain extent), and the number of frames is decreased while maintaining the scanning length (Figure 6). Overall, these findings suggest a generalization of these results to a considerable range of temporal resolution in the BOLD fMRI data.

### *4.8. Comparison with canonical metrics used to compare FCs*

With all the canonical methods of comparing FCs (e.g. Pearson's correlation coefficient, Euclidean distance), only the elements in the upper or lower triangular part of the FC are selected and vectorized. This means that regularization has no effect on those metrics since the regularization magnitude is added to the main diagonal which is ignored by all those metrics. It has already been shown by Venkatesh et al. (Venkatesh et al., 2020) that geodesic distance outperforms those metrics in uncovering individual fingerprint in FCs. They achieved this using a fixed non-optimal regularization magnitude ($\tau = 1$). Our results show that the combination of geodesic distance with an optimal regularization outperforms $\tau = 1$ identification rates, and hence, above mentioned canonical metrics by a considerable margin.

### *4.9 How to estimate the optimal regularization parameter and the resulting geodesic distances in a specific study*

We have observed that the optimal regularization that leads to maximum identification rates is dependent on the fMRI condition, brain parcellation, scanning length, and the number of frames. There might be other aspects of the data that influence such optimal value as well, such as voxel size. Hence, results suggest that when using geodesic distance to compare FCs, the regularization parameter must be estimated from the FC data of that study. Also, one should utilize sampling techniques to estimate a mean or median magnitude of regularization, along with the corresponding error. Once an appropriate regularization magnitude has been identified, one should regularize all FCs in the dataset by that amount, and then use geodesic distance to compare FCs. These steps have been tabulated for the benefit of the user of this framework (Table 4). It is important to remark that these resulting pairwise distances are better suited for establishing associations between functional connectivity and cognition, behavior, and neurological diseases at the individual level.

**Table 4:** A step by step outline of how to estimate and apply an optimal regularization magnitude (τ) to an FC dataset, such that individual fingerprint is maximized when using geodesic distances to compare FCs.

| Step 1 | Estimate test and retest FCs per subject from the fMRI data |
|---|---|
| Step 2 | For a wide range of regularization magnitude (τ): <br><br> a. Obtain a random sample of the FC dataset without replacement* <br> b. Regularize FCs by that regularization magnitude (τ) <br> c. Compute pairwise geodesic distances and obtain the identifiability matrix. <br> d. Estimate identification rate from the identifiability matrix <br><br> *Random samplings without replacement are performed to estimate mean behavior (and standard error) of identification rate with respect to the regularization |
| Step 3 | Identify the optimal regularization magnitude (τ*), such that (mean) identification rate is maximized |
| Step 4 | Regularize all FCs in the dataset by the optimal regularization magnitude (τ*) |
| Step 5 | Compare the optimally regularized FCs by using geodesic distance |
| Step 6 | (Optional) For every two subjects, average all 4 test/retest geodesic distances |
| Step 7 | Use those geodesic distances to establish associations with cognitive/clinical outcomes |

This process of estimating an optimal regularization from the data themselves, and then applying it back to the same data might seem biased, but we should emphasize that the optimal regularization is estimated to maximize individual fingerprint in the data and nothing else. It is not optimized for any group differences or for any neuro/psychiatric or behavioral score. The only desired output is maximal inter-individual differentiability so that the desired effects could be accurately captured at the individual-level.

It might also seem desirable to have a constant value of regularization (say 0.1 or 1) that is applicable to all datasets, without any considerable negative effects. But as we have observed, deviations from optimal regularization magnitudes could have detrimental effects on the measured individual fingerprint depending on a variety of factors. Hence, it is always recommended to estimate an optimal regularization magnitude from the data themselves, especially considering that it is extremely easy and computationally efficient to estimate.

*4.10. Limitations and future work*

One limitation of the geodesic distance, whether applied to regularized or unregularized FCs, is that it only provides a single numeric distance estimate between FCs and hence does not allow elementwise (or edgewise) analyses of the FCs (i.e. analysis focused on a particular brain region or a specific functional coupling between two brain regions). Although this limitation can be addressed by projecting FCs from the SPD manifold onto a tangent space of symmetric matrices, which would be Euclidean and allow the use of Euclidean algebra and calculus (Pervaiz et al., 2020; You & Park, 2021). Future work should explore these projections and how they interact with regularization magnitude.

One could also explore the effects of regularization on the identification rates when the test and retest sessions belong to different fMRI conditions (e.g. working-memory vs resting-state), analogous to (Finn et al., 2015) and (Venkatesh et al., 2020). To estimate the optimal amount of regularization based on functional connectivity fingerprint, one could go beyond test/retest of the same individual and assess identification rates when the test and retest sessions belong to twin-pairs (Monozygotic or Dizygotic). Finally, we can compare this straight forward main diagonal regularization with other kinds of regularization techniques that include off diagonal elements or add a variable amount to the elements of the main diagonal.

## 5. Conclusion

The use of the geodesic distance on full-rank or regularized rank-deficient FCs, has been shown to be a more principled and accurate method to compare FCs than canonical methods, ultimately leading to improved subject fingerprinting, as measured by identification rates. Here we combine geodesic distance with optimal regularization to uncover brain connectivity fingerprints by means of an incremental assessment of the magnitude of the regularization parameter. We show that optimal regularization that maximizes subject identification rates is highly dataset-dependent ─ it depends on the fMRI condition, on the brain parcellation used, scanning length, and on the number of frames used to compute the FCs.

## ACKNOWLEDGEMENTS

Data were provided [in part] by the Human Connectome Project, WU-Minn Consortium (Principle Investigators: David Van Essen and Kamil Ugurbil; 1U54MH091657) funded by the 16 NIH Institutes and Centers that support the NIH Blueprint for Neuroscience Research; and by the McDonnell Center for Systems Neuroscience at Washington University.

Previous versions of this article have appeared in a preprint posting on the Cornell University's arXiv.org server, under the Quantitative Biology category, with the sub-category of Neurons and Cognition. The preprint was first uploaded to the server on March 11$^{th}$ 2020, while a modified version was updated on August 29$^{th}$ 2020.

## FUNDING INFORMATION

JG acknowledges financial support from NIH R01EB022574, NIH R01MH108467, Indiana Alcohol Research Center P60AA07611, and Purdue Discovery Park Data Science Award "Fingerprints of the Human Brain: A Data Science Perspective". JH has been partially supported by the grant NIH R01MH108467. EA acknowledges financial support from the SNSF Ambizione project "Fingerprinting the brain: network science to extract features of cognition, behavior and dysfunction" (grant number PZ00P2_185716). LP's research has been supported by the National Institute of Mental Health (R01 MH071589 and R01 MH112517).